# Nucleation and propagation of excitation fronts in self-excited systems

I.B. Shiroky[1,*], A. Papangelo[2,**], N. Hoffmann[2], O.V. Gendelman[1]

1Technion—Israel Institute of Technology, Faculty of Mechanical Engineering, Technion City, Haifa, 32000, Israel

2Hamburg University of Technology, Department of Mechanical Engineering, Am Schwarzenberg-Campus 1, 21073, Hamburg, Germany

email: (*) shiroky@campus.technion.ac.il, (**) antonio.papangelo@poliba.it

**Abstract:** Frictional interfaces exhibits a very rich dynamical behavior due to frictional interactions. This work focuses on the transition between spatially *localized* and *propagating* stick-slip motion. To overcome the difficulties related to the non-smoothness of the friction law we have studied two models: (model I) a friction-excited chain of weakly coupled oscillators excited by a frictional moving belt and (model II) a chain of Van der Pol oscillators. The two models show very similar dynamical features. Both exhibit discrete breathers (i.e. spatially localized periodic solutions) solutions for low coupling and show propagation along the chain of high amplitude limit cycles for stronger elastic coupling between the unit cells. In both models the transition from discrete breathers to propagating limit cycles happens through a nucleation process and, above a certain critical coupling, the velocity of propagation scales linearly with the elastic coupling coefficient. We suggest that dynamical features similar to what was demonstrated in this work are expected for models that are different quantitatively but preserve the same single-unit topology.

1. Introduction

Tribology is an area of very active research nowadays as for the several application fields where it plays a fundamental role [1]. Although the basics of frictional interaction are well known since ancient times [2], describing frictional interactions is still challenging nowadays. Interfaces have revealed to be dynamic entities, as the exceptional experiments from Fineberg and co-authors have clearly shown [3, 4, 5]. Transition to sliding is mediated by interfacial rupture fronts which transverse the interface with velocity spanning roughly two orders of magnitude from "slow" to supershear velocity [3].

Difficulties in modelling frictional interfaces come from their intrinsic multiscale nature, which involves a wide spectrum of both length- and time-scales [1]. Frictional interactions are at the basis of the developing of self-excited vibrations, usually referred as Friction-Induced Vibrations (FIV), which represent a serious problem in different engineering applications, such as machine tool/work-piece interaction, brake systems, wheel/rail contact [6, 7, 8]. Although the basic triggering mechanisms have been studied for decades, avoiding them, and the associated noise, is still an arduous task. Difficulties in studying the emergence of FIV [9, 10, 11, 12] are related to the nonlinear interactions that take place at rough interfaces, involving nonlinear contact stiffness [13] and non-smooth friction laws [14]. This gives rise to a wide spectrum of possible dynamical states, which range from regular/irregular motion [15, 16] to localized/propagating interfacial stick-slip vibrations [17, 18]. The emergence of these dynamical states is still poorly understood. Papangelo et al. [18] have shown that weakly coupled chains of frictional oscillators may experience spatially localized stick-slip vibrations. In their model different localized patterns arose, depending on the initial conditions, with periodic and irregular stick-slip motion. They ascribed this multiplicity of solutions to the inherent bi-stability of the "unit cell" that composed the chain (i.e. the isolated friction-excited oscillator), which, for certain range of substrate velocity, experienced a subcritical Hopf-bifurcation [19, 18]. Besides the localized states Papangelo et al. [18] have pointed to the appearance of a parametric region where stick-slip oscillations propagates along the chain. These propagating stick-slip states have been observed also by Costagliola at al. [20], who extended the interfacial spring-block chain to two dimensions [21, 22].

The aim of this paper is to provide a better understanding of the transition between spatially *localized* and *propagating* stick-slip motions. Nevertheless, due to the non-smooth interactions, it is difficult to obtain analytical approximation of the dynamical states in friction-excited systems. To overcome this difficulty, we will adopt two different models which share common features: in the first part of the paper a chain of elastically nearest-neighbor coupled friction-excited oscillators in contact with a moving substrate will be studied, while, in the second part of the paper, we will address a smooth model constituted by a chain of coupled Van der Pol oscillators (model II). It will turn out that the two models share many common features, implying on possible generic properties. Both models have stable regimes of localized solutions and travelling waves. Also, in both cases, the transition between the localized solution and the propagating one, occurs through a nucleation process, which will be explained in details. The final similarity is that the velocity of propagation can be nearly linearly scaled as a function of the coupling coefficient.

1. Model I

1.1 A Friction-excited chain of weakly coupled oscillators

Let us consider a chain of $N_{dof}$ oscillators which are elastically coupled to the nearest-neighbor via a linear spring $c$. Each oscillator has mass $m$, ground stiffness $k$, viscous damping coefficient $d$ and is pressed by a constant normal force $F_N$ against a substrate that moves at a certain driving velocity $v_d$. The equilibrium equation for the $n^{th}$ oscillator reads

$$my''_n + dy'_n + ky_n + \eta_c(2y_n - y_{n+1} - y_{n-1}) = F(v_{rel,n}) \qquad (1)$$

where $y_n(t)$ defines the oscillator position at time $t$, a prime indicates the derivative with respect to time, $v_{rel,n} = y'_n - v_d$ is the relative velocity and $F(v_{rel,n})$ accounts for the rate dependent friction force exchanged at the interface between the $n^{th}$ oscillator and the substrate. We assume free-free boundary conditions, hence the last oscillator in the chain is not coupled with the first. Assuming that the friction force decays exponentially with the relative velocity $v_{rel}$ we write:

$$\mu(v_{rel}) = \mu_{st} + (\mu_{st} - \mu_d)e^{\left(-\frac{|v_{rel}|}{v_0}\right)} \qquad (2)$$

where $\mu_{st}$ is the static friction coefficient, $\mu_d$ is the dynamic friction coefficient with $\mu_{st} > \mu_d$, $v_0$ is a reference velocity and from equation (2) it follows that $\mu(0) = \mu_{st}$ and $\mu(v_{rel} \to \infty) = \mu_d$. Hence, the friction force can be written as:

$$\begin{cases} F = -\text{sgn}(v_{rel}) F_N \mu(v_{rel}) & v_{rel} \neq 0 \\ |F| < \mu_{st} F_N & v_{rel} = 0 \end{cases} \qquad (3)$$

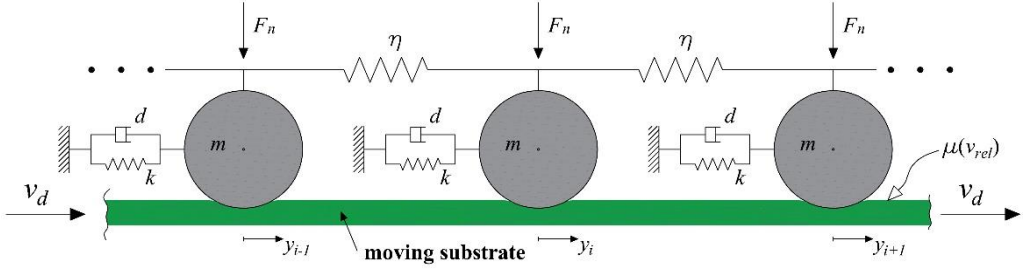

*Figure 1 - Friction-excited chain of elastically coupled oscillators. The friction force between each oscillator and the substrate is modeled by eq.(3).*

The following non-dimensional parameters are introduced

$$\omega_n = \sqrt{\frac{k}{m}}, \quad y_0 = \frac{F_N}{k}, \quad c = \frac{\eta_c}{k}, \quad \xi = \frac{d}{2\sqrt{dm}}, \quad \tau = \omega_n t \qquad (4)$$

The dimensionless oscillator position is $\tilde{y} = y/y_0$. Substituting in eq. (1) $\frac{d}{dt}$ with $\omega_n \frac{d}{d\tau}$ the equilibrium equation is written in dimensionless form

$$\ddot{\tilde{y}}_n + 2\xi \dot{\tilde{y}}_n + \tilde{y}_n + c(2\tilde{y}_n - \tilde{y}_{n+1} - \tilde{y}_{n-1}) = \tilde{F}(\tilde{v}_{rel,n}) \qquad (5)$$

where $\tilde{v}_{rel} = \frac{v_{rel}}{\omega_n x_0}$, a dot superposed indicates $d/d\tau$ and the dimensionless friction force $\tilde{F}$ reads as:

$$\begin{cases} \tilde{F} = -\text{sgn}(\tilde{v}_{rel}) \mu(\tilde{v}_{rel}) & \tilde{v}_{rel} \neq 0 \\ |\tilde{F}| < \mu_{st} & \tilde{v}_{rel} = 0 \end{cases} \qquad (6)$$

Given the initial conditions $y_n(\tau = 0) = y_{n,0}$ and $\dot{y}_n(\tau = 0) = \dot{y}_{n,0}$ ($n \in [1, N_{dof}]$) the equations of motion can be integrated in time. We used MATLAB built-in ode function "ode23t" which adopts the trapezoidal rule without numerical damping. The non-smooth friction force is implemented using the switch-model as proposed by Leine et al. [23], which defines a narrowband of vanishing relative velocity where the stick equations are solved. We assumed that the oscillator sticks to the substrate for $|\tilde{v}_{rel}| < 10^{-4}$.

### 1.2 Localized and propagating stick-slip limit cycles

Before we look at the chain response we recall that Papangelo et al. [19] have previously studied the response of the isolated single degree of freedom (in the following the "unit cell" that composes the chain in Figure 1) showing that there exist three regions with different dynamics.

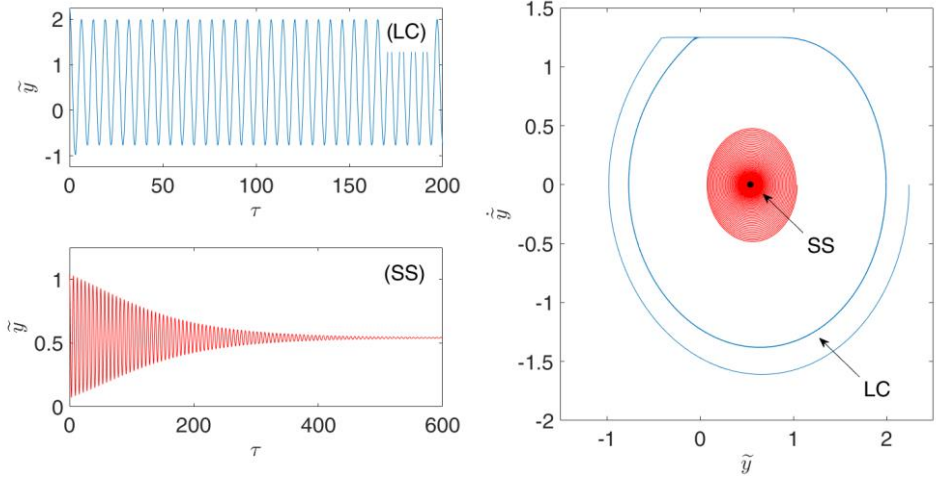

*Figure 2 - Right panel: state space plot of the two possible states for the isolated single degree of freedom, i.e. the "unit cell" of the chain, (i) Limit Cycle (LC) (ii) Steady Sliding (SS). The corresponding time series are plotted in the left panels. In the simulations we used the same parameters in both simulations $\mu_{st}=1, \mu_d=0.5, \tilde{v}_0=0.5, \xi=0.05, \tilde{v}_d=1.25,$ except for the initial conditions.*

For low substrate velocity high amplitude stick-slip limit cycle are sustained, while for high substrate velocity Steady Sliding (SS) is stable. In the transition from low to high velocity a region of bi-stability exist, where both stick-slip limit cycles and SS are stable and the system dynamics is determined by the initial conditions. In Figure 2 the right panel shows the system dynamics in the state space representation for the parameters reported in (7), $\tilde{v}_d=1.25$ and differing initial conditions. One easily recognizes that both states SS and stick-slip LC are stable for this parameters set (see also Figure 2 left panels where the displacement evolution $\tilde{y}(t)$ with time is reported for both cases (SS and stick-slip LC)).

For the chain of friction-excited oscillators described by the set of equations (1) we assume the following set of parameters:

$$\mu_{st}=1; \quad \mu_d=0.5; \quad \tilde{v}_0=0.5; \quad \xi=0.05; \tag{7}$$

In the following we will name "excited oscillators" those which initial conditions are selected close to stick-slip limit cycle of the isolated unit cell, while "not excited oscillators" correspond to those which initial conditions lie in the neighbor of the unit cell SS solution. We will assume $\dot{y}_{n,0}=0$ for ($n \in [1, N_{dof}]$).

Let us consider a chain of $N_{dof}=31$ oscillators. In Figure 3a we report the vibration amplitude $\tilde{Y}_n$ for $c=0.4$ % for a simulation where three oscillators ($N_{size}=3$, "nucleus size") were excited at $\tau=0$ (the numbers $15-16-17$ in the chain). Notice that $\tilde{Y}_n$ is shown for $\tau=[200,400,600]$ but it is not possible to distinguish the three curves as they are superimposed. In Figure 3b $\tilde{Y}_n$ is plotted versus the oscillator number $n$ for a simulation where we excited only the first oscillator on the edge of the chain. In Figure 3b the elastic coupling coefficient is $c=2$ % and we plotted $\tilde{Y}_n$ for $\tau=[400,700,1000]$. Clearly, changing the initial conditions and depending on the elastic coupling strength, different dynamical behaviors are observed: Figure 3a shows a Discrete Breather (DB) solution, i.e. high amplitude stick-slip vibration remains spatially localized and surrounded by a Steady-Sliding (SS) domain (with vanishing amplitude). Instead, the stronger coupling in Figure 3b promotes propagation of the excited state into the steady sliding domain.

### 1.3 Dependence on nucleus size, substrate velocity and elastic coupling

In this section we study how the system dynamics, particularly the transition from DB to propagating stick-slip limit cycles (Front Propagation, FP) depends on nucleus size $N_{size}$, substrate velocity $\tilde{v}_d$ and elastic coupling $c$. We perform a series of numerical investigations for a chain with $N_{dof} = 31$ oscillators and $\tilde{v}_d = 1.25$ to study the effect that the nucleus dimension $N_{size}$ has on the critical elastic coupling $(c)_{crit}$ at which propagation of stick-slip limit cycles is triggered. For the parameters given in (7) we varied the dimension of the excited nucleus $N_{size} = [1,3,5,7,9]$, which is placed in the middle of the chain (Figure 4, dashed red line, "bulk" excitation) or on the edge of the chain (Figure 4, black solid line, "edge" excitation). Figure 4 shows that in general a stronger coupling promotes propagation of high amplitude vibrations along the chain. Nevertheless for "edge" excitation $(c)_{crit}$ drops of about 40% moving from $N_{size} = 1$ to $N_{size} > 1$, while for "bulk" excitation $(c)_{crit}$ remains almost insensitive to $N_{size}$.

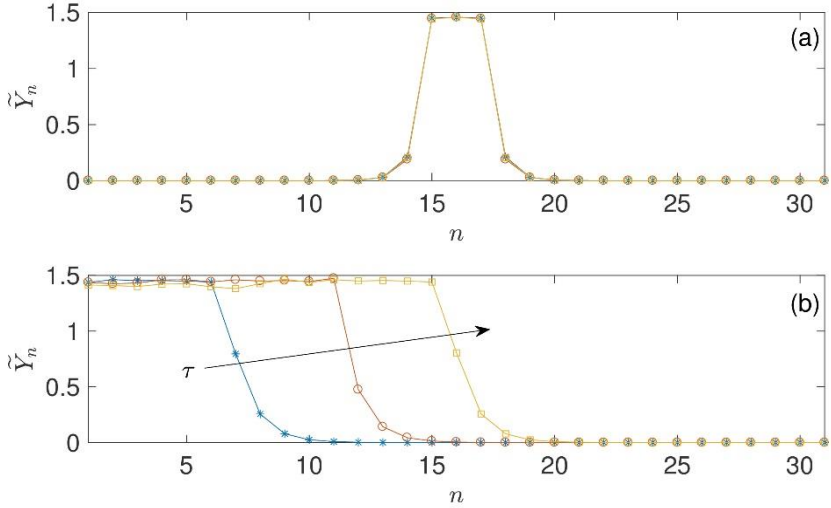

Figure 3 - For both panels: vibration amplitude envelope $\tilde{Y}_n$ with respect to the oscillator number $n$ for a numerical simulation with $N_{dof} = 31$ and $\tilde{v}_d = 1.25$. The rest of the parameters used are given in (7). (a) Discrete breather solution. Excited oscillator numbers $[15 - 16 - 17]$, i.e. nucleus of size $N_{size} = 3$, $c = 0.4\%$. The vibration amplitude is plotted for $\tau = [200, 400, 600]$. (b) Propagating stick-slip solution. The first oscillator on the edge of the chain is excited, $c = 2\%$ and the vibration amplitude is plotted for $\tau = [400, 700, 1000]$.

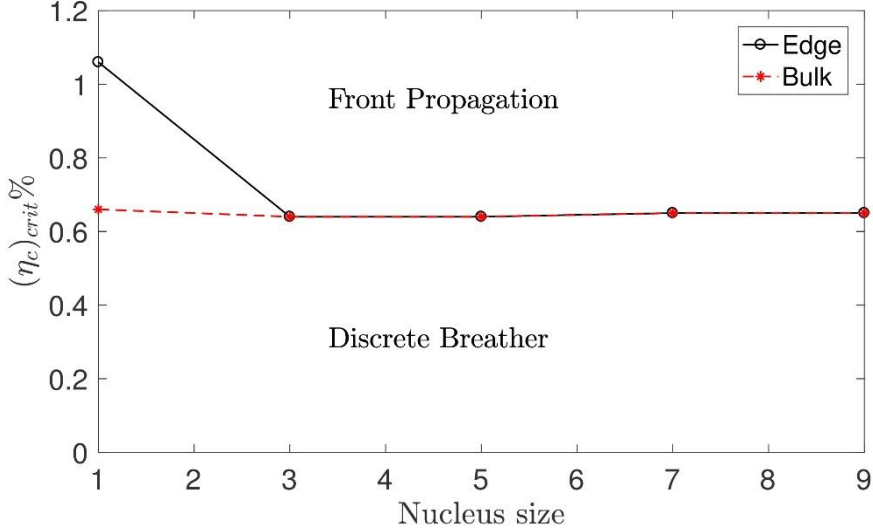

*Figure 4 - The critical elastic coupling coefficient $(c)_{crit}$ as a function of the nucleus size $N_{size}$ is plotted for both "edge" and "bulk" excitation. Below (above) the critical value discrete breather (front propagation) are expected. For edge excitation (black solid line, the chain is excited from one side) the critical stiffness for propagation falls of about 40% moving from $N_{size} = 1$ to $N_{size} > 1$. In case of bulk excitation (the excited oscillators lie in the middle of the chain) the critical elastic coupling is almost insensitive to $N_{size}$.*

We look now for the parametric regions where the different dynamics have to be expected depending on the elastic coupling $c$ and on the driving velocity $\tilde{v}_d$. For the considered friction-excited chain, we assumed $N_{dof} = 31$, fixed $N_{size} = 1$ ("bulk" excitation is used, i.e. the nucleus is excited in the middle of the chain) and computed a series of time marching numerical simulations for $c \in [0,0.1]$ and $\tilde{v}_d \in [1, 2]$, the remaining parameters being those in (7). Figure 5 shows the dynamical states corresponding to each couple $(c, \tilde{v}_d)$, in particular blue triangles, red circles, green squares stay respectively for Front Propagation (FP), Discrete Breathers (DB) and Steady-Sliding (SS). The map (Figure 5) shows that DB solutions exist only for low elastic coupling, as it should be expected. Increasing the coupling strength the DB region shrinks and only the two homogenous states remain. Indeed, for finite size chains, after propagation all the oscillators experience stick-slip limit cycles, or, for higher velocity steady-sliding with no vibrations.

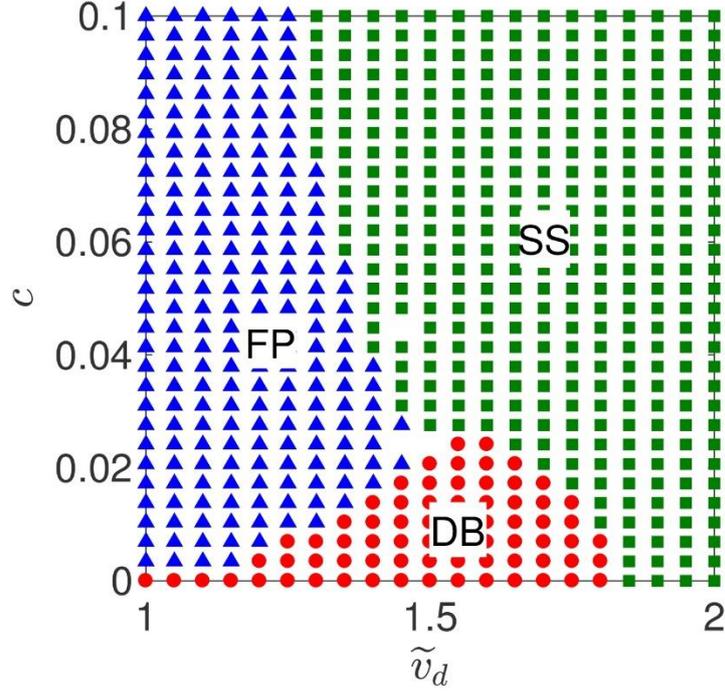

*Figure 5 - Map of the possible dynamical states in the plane $(c, \tilde{v}_d)$. The system has $N_{dof} = 31$, and we fixed $N_{size} = 1$ ("bulk" excitation). Front Propagation (FP), Discrete Breathers (DB) and Steady-Sliding (SS) solutions correspond respectively to the blue triangles, red circles and green squares. $N_{dof} = 31$, $N_{size} = 1$ ("bulk" excitation) and the remaining parameters are those in (7).*

Lastly, we look at the velocity of propagation $\tilde{v}_{front}$ of high amplitude stick-slip limit cycle as a function of the coupling strength $c$. Figure 6 shows that, after a critical coupling strength is reached, the velocity of propagation $\tilde{v}_{front}$ rapidly increases and then scales linearly with $c$. All the numerical experiments in Figure 6 have $N_{size} = 1$ ("edge" excitation), $N_{dof} = 31$ and $\tilde{v}_d = 1.25$.

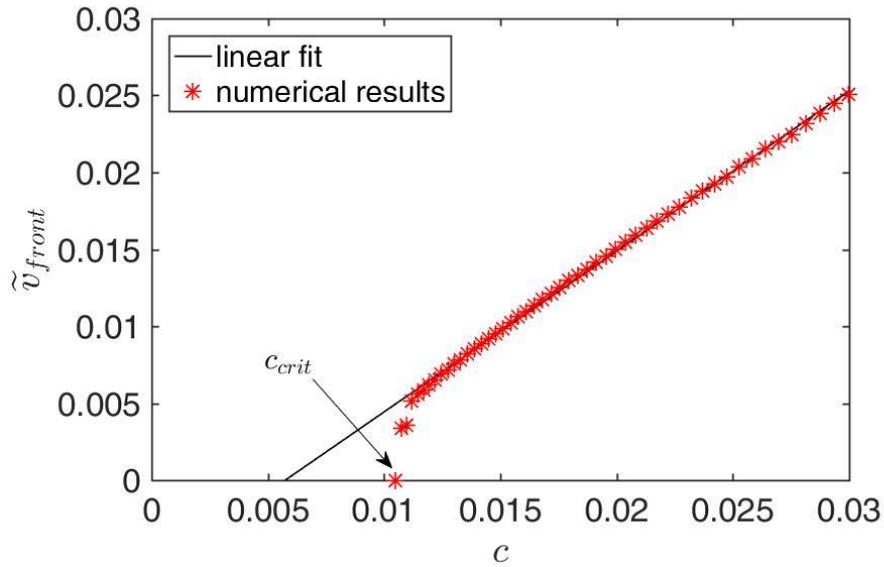

*Figure 6 - Velocity of propagation of stick-slip limit cycle as a function of c. For $c < c_{crit}$ there is no propagation. For $c > c_{crit}$ the velocity of propagation of stick-slip limit cycles scales linearly with the elastic coupling coefficient c (except in the vicinity of $c_{crit}$). $N_{size} = 1$ ("edge" excitation), $N_{dof} = 31$ and $\tilde{v}_d = 1.25$.*

2. Model II: A chain of coupled Van der Pol (VdP) oscillators

In this section we address a system of coupled VdP oscillators (8). The reason for this selection is that the isolated VdP oscillator is probably the simplest available model in the family of systems that are bi-stable, in a sense of having two dynamic equilibrium states: a limit cycle and a fixed point. These topological characteristics are similar to Model I and thus, in a sense, the current model is a variation of the previous one. The smoothness of Model II allows application of several analytical techniques, which might lead to a deeper insight on Model I as well, due to possible robustness of the topologically similar models.

This system was previously explored in [24]. The system is written as follows:

$$\ddot{y}_n + y_n + \varepsilon c(2y_n - y_{n-1} - y_{n+1}) + \varepsilon \dot{y}_n(1 - ay_n^2 + by_n^4) = 0 \quad (8)$$

where $\varepsilon \ll 1$ is a small parameter.

2.1 Approximate solution for the discrete breather

First solution to explore is the discrete breather (a localized solution). It is natural to look for this type of solutions in the low coupling range. A numerical example is brought in Figure 7. The obtained breather is instantiated by originally placing a single particle close to the stable limit cycle (calculated for the SDOF model). The result is a classically localized solution with an exponential decay. In terms of this paper we call this solution: a discrete breather with a nucleus of size 1. Nuclei of other sizes and the significance of this issue will be addressed later.

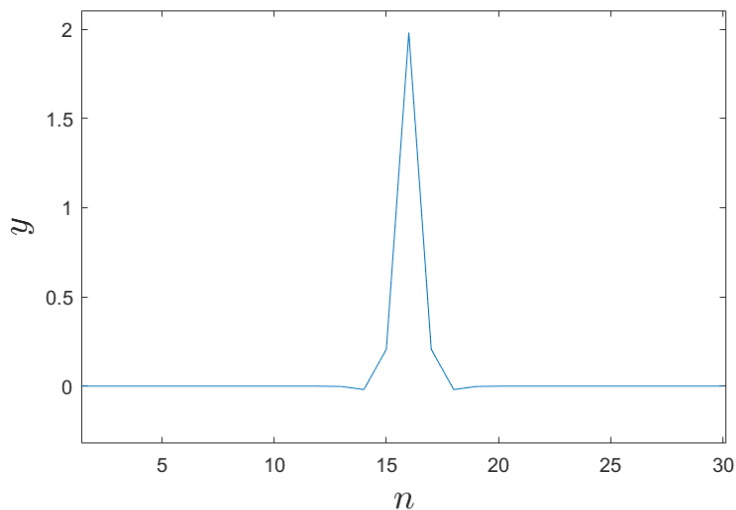

*Figure 7 – A discrete breather with a single particle at the stable limit cycle (nucleus of size 1). Parameters:* $c = 0.1$, $a = 2.5$, $b = 0.5$, $\varepsilon = 0.01$

We are looking for a solution of the following form:

$$y_n = \alpha_0 e^{k_1 |n|} \sin(\omega t + k_2 |n|) \tag{9}$$

Where $\alpha_0$ is the unknown amplitude of the central particle ($n=0$) which is undergoing a periodic oscillation within the stable limit cycle. $k_1$ is the rate of exponential decay of the localized solution, $k_2$ is the phase difference between the particles. We assume constant $k_1$ and $k_2$.

The steps to derive the approximate solution are as follows. First, in 2.1.1 we perform a multiple scales analysis for the central particle ($n=0$). Then, in 2.1.2, we consider an exponential decay for the rest of the chain ($n \neq 0$). Finally, in 2.1.3 we solve the set of Eqs. achieved in 2.1.1 and 2.1.2 and verify the results numerically.

### 2.1.1 Multiple scales analysis for the central particle

We address the central particle ($n=0$) in the set of Eqs. (8). Also, we exploit the symmetry of the localized solution $y_1 = y_{-1}$. Hence, one achieves:

$$\ddot{y}_0 + y_0 + 2\varepsilon c(y_0 - y_1) + \varepsilon \dot{y}_0 (1 - a y_0^2 + b y_0^4) = 0 \tag{10}$$

Obviously Eq. (10) has two unknown functions $y_0, y_1$. By substitution of the assumed form of the solution (9) into (10) one achieves:

$$\ddot{y}_0 + y_0 + 2\varepsilon c(y_0 - \alpha_0 e^{k_1} \sin(\omega t + k_2)) + \varepsilon \dot{y}_0 (1 - a y_0^2 + b y_0^4) = 0 \tag{11}$$

Eq. (11) is an ode (ordinary differential equation) and the approximate solution is obtained by means of a multiple scales analysis similarly to a previous work of Rand [25].

The frequency of oscillations is expanded in a series with respect to $\varepsilon$:

$$\omega = 1 + \varepsilon \omega_1 + \varepsilon^2 \omega_2 + \ldots \tag{12}$$

Let us introduce the following time scales:

$$\hat{t} = \omega t, \quad \tau = \varepsilon t \tag{13}$$

The derivatives of $y_0$ with respect to the new time scales are given by:

$$\begin{aligned} \dot{y}_0 &= \omega y_{\hat{t}} + \varepsilon y_\tau \\ \ddot{y}_0 &= \omega^2 y_{\hat{t}\hat{t}} + 2\omega\varepsilon y_{\tau \hat{t}} + \varepsilon^2 y_{\tau\tau} \end{aligned} \tag{14}$$

Substitution of (14) into (10) and omitting $O(\varepsilon^2)$ yields:

$$\omega^2 y_{\hat{t}\hat{t}} + 2\omega\varepsilon y_{\hat{t}\tau} + y_0 + 2\varepsilon c(y_0 - \alpha_0 e^{k_1} \sin(\hat{t} + k_2)) + \varepsilon \omega y_{\hat{t}} (1 - a y_0^2 + b y_0^4) = 0 \tag{15}$$

Expression (12) is substituted into (15) and $O(\varepsilon^2)$ terms are omitted:

$$(1+2\varepsilon\omega_1)y_{0,\hat{t}\hat{t}} + 2\varepsilon y_{0,\hat{t}\tau} + y_0 + 2\varepsilon c\left(y_0 - \alpha_0 e^{k_1}\sin(\hat{t}+k_2)\right) + \varepsilon y_{\hat{t}}\left(1 - ay_0^2 + by_0^4\right) = 0 \qquad (16)$$

Then, $y_0$ is expanded in a power series with respect to $\varepsilon$:

$$y_0 = y_0^{(0)} + \varepsilon y_0^{(1)} + \ldots \qquad (17)$$

By substituting of (17) into (16) and collecting similar powers of $\varepsilon$ one achieves the following set of equations:

$$y_{0,\hat{t}\hat{t}}^{(0)} + y_0^{(0)} = 0 \qquad (18)$$

$$y_{0,\hat{t}\hat{t}}^{(1)} + y_0^{(1)} = -2\omega_1 y_{0,\hat{t}\hat{t}}^{(0)} - 2y_{0,\hat{t}\tau}^{(0)} - 2c\left(y_0^{(0)} - \alpha_0 e^{k_1}\sin(\hat{t}+k_2)\right) - y_{0,\hat{t}}^{(0)}\left(1 - a\left(y_0^{(0)}\right)^2 + b\left(y_0^{(0)}\right)^4\right) \qquad (19)$$

The solution of (18) is:

$$y_0^{(0)} = A(\tau)e^{i\hat{t}} + A(\tau)^* e^{-i\hat{t}} \qquad (20)$$

Where $(*)$ depicts a complex conjugate. We substitute (20), $\sin(\hat{t}+k_2) = \dfrac{e^{ik_2}e^{i\hat{t}} - e^{-ik_2}e^{-i\hat{t}}}{2i}$ into (19), collect the coefficient of $e^{i\hat{t}}$ and demand equality to $0$ to avoid secular terms. $' \equiv \dfrac{d}{d\tau}$

$$2iA|A|^4 b - iA|A|^2 a + iA + 2cA + 2iA' - 2A\omega_1 + i\alpha_0 ce^{k_1}e^{ik_2} = 0 \qquad (21)$$

Complex amplitude $A$ can be presented in a polar form:

$$A = R \cdot e^{i\theta} \qquad (22)$$

Substitution of (22) into (21) and division by $e^{i\theta}$ leads to:

$$2iR^5 b - iR^3 a + iR + 2cR + 2iR' - 2\theta' R - 2\omega_1 R + i\alpha_0 ce^{k_1}e^{i(k_2-\theta)} = 0 \qquad (23)$$

Separation into real and imaginary parts yields:

$$\begin{cases} 2R^5 b - R^3 a + R + 2R' + \alpha_0 ce^{k_1}\cos(k_2-\theta) = 0 \\ 2cR - 2\theta' R - \alpha_0 ce^{k_1}\sin(k_2-\theta) - 2\omega_1 R = 0 \end{cases} \qquad (24)$$

We are interested in a steady state solution with a constant amplitude, $R'=0, \theta'=0$:

$$\begin{cases} 2R^5 b - R^3 a + R + \alpha_0 ce^{k_1}\cos(k_2-\theta) = 0 \\ 2cR - \alpha_0 ce^{k_1}\sin(k_2-\theta) - 2\omega_1 R = 0 \end{cases} \qquad (25)$$

For self-consistency we demand $\alpha_0 = 2R$ and we achieve:

$$4c^2 e^{2k_1} = \left(-2R^4 b + R^2 a - 1\right)^2 + 4\left(c - \omega_1\right)^2 \tag{26}$$

Eq. (26) has three unknowns: $R, k_1, \omega_1$. An extra relation is required to remain with two unknowns (in the next subsection another equation with two unknowns will be derived and will allow to solve the system uniquely). Hence, we make the following assumption:

$$\omega_1 = c + \delta; \quad \delta \ll c \tag{27}$$

By substitution of (27) into (26), we achieve that:

$$4c^2 e^{2k_1} \approx \left(-2R^4 b + R^2 a - 1\right)^2 \tag{28}$$

Of course, assumption (27) has to verified for self-consistency through numerical simulations.

To conclude this subsection, we obtain the following relations:

$$e^{k_1} = \frac{-\alpha_0^4 b + 2\alpha_0^2 a - 8}{16c} \tag{29}$$

$$\omega = 1 + \varepsilon c \tag{30}$$

### 2.1.2 The balance of the linear sub-structure

In this subsection we demand an exponential decay according to (9). We introduce a complex variable $\Psi_n$, and express $y_n$ and its derivatives through $\Psi_n$ in the following way:

$$y_n = -i\frac{\Psi_n - \Psi_n^*}{2\omega}, \quad \dot{y}_n = \frac{\Psi_n + \Psi_n^*}{2}, \quad \ddot{y}_n = \dot{\Psi}_n - i\omega\left(\frac{\Psi_n + \Psi_n^*}{2}\right) \tag{31}$$

Substitution of (31) into (8) yields:

$$\begin{aligned}&\dot{\Psi}_n - i\omega\left(\frac{\Psi_n + \Psi_n^*}{2}\right) - i(1 + 2\varepsilon c)\left(\frac{\Psi_n - \Psi_n^*}{2\omega}\right) \\ &+ \frac{ic\varepsilon}{2\omega}\left(\Psi_{n-1} - \Psi_{n-1}^* + \Psi_{n+1} - \Psi_{n+1}^*\right) + \varepsilon\frac{\Psi_n + \Psi_n^*}{2}\left(1 + a\left(\frac{\Psi_n - \Psi_n^*}{2\omega}\right)^2 + b\left(\frac{\Psi_n - \Psi_n^*}{2\omega}\right)^4\right)\end{aligned} \tag{32}$$

By substitution of $\Psi_n = \varphi_n e^{i\omega t}$ into (32) and by dividing the equation by $e^{i\omega t}$ we achieve:

$$\begin{aligned}&\dot{\varphi}_n + \frac{i\omega}{2}\varphi_n - \frac{i\omega \varphi_n^* e^{-2i\omega t}}{2} - i(1 + 2\varepsilon c)\left(\frac{\varphi_n - \varphi_n^* e^{-2i\omega t}}{2\omega}\right) \\ &+ \frac{i\varepsilon c}{2\omega}\left(\varphi_{n-1} - \varphi_{n-1}^* e^{-2i\omega t} + \varphi_{n+1} - \varphi_{n+1}^* e^{-2i\omega t}\right) \\ &+ \varepsilon\frac{\varphi_n + \varphi_n^* e^{-2i\omega t}}{2}\left(\begin{aligned}&1 + a\frac{\varphi_n^2 e^{2i\omega t} - 2|\varphi_n|^2 - \left(\varphi_n^*\right)^2 e^{-2i\omega t}}{4\omega^2} \\ &+ b\frac{\varphi_n^4 e^{4i\omega t} - 4|\varphi_n|^2 \varphi_n^2 e^{2i\omega t} + 6|\varphi_n|^4 - 4|\varphi_n|^2 \left(\varphi_n^*\right)^2 e^{-2i\omega t} + \left(\varphi_n^*\right)^4 e^{-4i\omega t}}{16\omega^4}\end{aligned}\right) = 0\end{aligned} \tag{33}$$

Averaging of Eq. (33) leads to the following expression envelope of the solution:

$$\dot{\varphi}_n + i\left(\omega - \frac{1}{\omega}\right)\frac{\varphi_n}{2} - \frac{i\varepsilon c}{2\omega}(2\varphi_n - \varphi_{n-1} - \varphi_{n+1}) + \varepsilon\frac{\varphi_n}{2}\left(1 - a\frac{|\varphi_n|^2}{4\omega^2} + b\frac{|\varphi_n|^4}{8\omega^4}\right) = 0 \quad (34)$$

We propose the following solution form in steady state $\varphi_n = \alpha_0 e^{(k_1 + ik_2)n}$, where $\alpha_0, k_1, k_2$ are constants. Substitution into (34) yields:

$$\frac{i\omega}{2} - i(1+2\varepsilon c)\left(\frac{1}{2\omega}\right)$$
$$+ \frac{i\varepsilon c}{2\omega}\left(e^{-(k_1+ik_2)} + e^{(k_1+ik_2)}\right) + \varepsilon\left(\frac{1}{2} - a\frac{\alpha_0^2 e^{2k_1 n}}{8\omega^2} + b\frac{\alpha_0^4 e^{4k_1 n}}{16\omega^4}\right) = 0 \quad (35)$$

Separation of Eq. (35) into a set of 2 real equations and substitution of (30) leads to:

$$\sin k_2 = \frac{2(1+\varepsilon c)}{c(e^{k_1} - e^{-k_1})}\left(\frac{1}{2} - a\frac{\alpha_0^2 e^{2k_1 n}}{8(1+\varepsilon c)^2} + b\frac{\alpha_0^4 e^{4k_1 n}}{16(1+\varepsilon c)^4}\right)$$
$$\cos k_2 = -\frac{\varepsilon c}{(e^{k_1} + e^{-k_1})} \quad (36)$$

In the limit $\varepsilon \to 0$ a second relation between $k_1$ and $\alpha_0$ is obtained:

$$\frac{2}{c(e^{k_1} - e^{-k_1})}\left(\frac{1}{2} - a\frac{\alpha_0^2 e^{2k_1 n}}{8} + b\frac{\alpha_0^4 e^{4k_1 n}}{16}\right) = -1, \quad k_2 = -\frac{\pi}{2} \quad (37)$$

### 2.1.3 Verification of the approximate solution

From Eqs. (29), (37) it is possible to (numerically) extract the values of $\alpha_0$ and $k_1$. Then, this allows the estimation of the amplitudes of all particles according by the simple expression (38). Numerical verification for the particles $n=0, n=1$ is presented in Figure 8 and a very good match is observed.

$$A_n = \alpha_0 e^{k_1|n|} \quad (38)$$

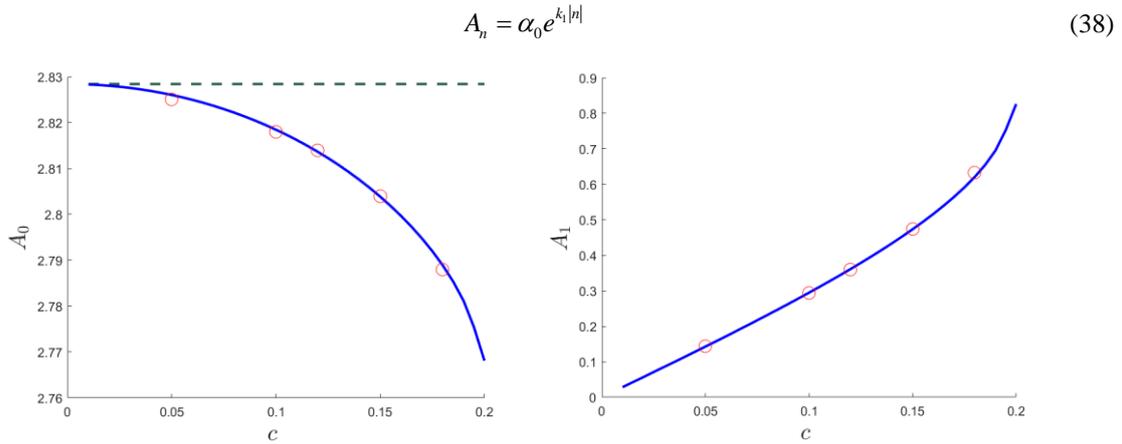

Figure 8 - Amplitude of particles $n=0, n=1$ as a function of coupling $c$. Solid line – analytical, circles – numerical, Parameters: $a = 2.5, b = 0.5, \varepsilon = 0.01$

### 2.1.4 Stability analysis

Here we examine the stability of discrete breather solution developed in the previous section. Let us define the state vector of the system:

$$X = \begin{pmatrix} Y \\ V \end{pmatrix}, \quad Y = \begin{pmatrix} y_1 \\ y_2 \\ \vdots \\ y_N \end{pmatrix}, \quad V = \dot{Y} \tag{39}$$

System (8) is then rewritten in the following way:

$$\dot{X} = f(X) = \begin{pmatrix} x_{N+1} \\ x_{N+2} \\ \vdots \\ -x_1 - \varepsilon c(2x_1 - x_2 - x_N) - \varepsilon x_{N+1}(1 - ax_1^2 + bx_1^4) \\ -x_2 - \varepsilon c(2x_2 - x_3 - x_1) - \varepsilon x_{N+2}(1 - ax_2^2 + bx_2^4) \\ \vdots \\ -x_N - \varepsilon c(2x_N - x_1 - x_{N-1}) - \varepsilon x_{2N}(1 - ax_N^2 + bx_N^4) \end{pmatrix} \tag{40}$$

Let $\hat{X}$ a solution of system (8); $\dot{\hat{X}} = f(\hat{X})$. Let $X = \hat{X} + \delta$, a perturbation of the solution. Then,

$$\dot{\hat{X}} + \dot{\delta} = J\big|_{X=\hat{X}} \delta + f(\hat{X}) + O(\delta^2) \quad \rightarrow \quad \dot{\delta} = J\big|_{X=\hat{X}} \delta \tag{41}$$

Where J is a $2N \times 2N$ Jacobian matrix expressed in the following way:

$$J = \begin{pmatrix} 0_{N \times N} & I_{N \times N} \\ A_{N \times N} & B_{N \times N} \end{pmatrix}$$

$$A_{N \times N} = \begin{bmatrix} -1 - 2\varepsilon c - \varepsilon x_{N+1}(1 - 2ax_1 + 4bx_1^3) & \varepsilon c & 0 & \varepsilon c \\ \varepsilon c & -1 - 2\varepsilon c - \varepsilon x_{N+2}(1 - 2ax_2 + 4bx_2^3) & \ddots & 0 \\ 0 & \ddots & \ddots & \varepsilon c \\ \varepsilon c & 0 & \varepsilon c & -1 - 2\varepsilon c - \varepsilon x_{2N}(1 - 2ax_N + 4bx_N^3) \end{bmatrix}$$

$$B_{N \times N} = \begin{bmatrix} -\varepsilon(1 - ax_1^2 + bx_1^4) & 0 & 0 \\ 0 & -\varepsilon(1 - ax_2^2 + bx_2^4) & \\ & & 0 \\ 0 & & -\varepsilon(1 - ax_N^2 + bx_N^4) \end{bmatrix}$$

(42)

We evaluate the monodromy matrix numerically, by using the approximate solution obtained in section 2.1. Then we evaluate the eigenvalues of the matrix, also known as "Floquet multipliers". According to the theory, loss of stability corresponds to escape of multipliers from the unit circle. In Figure 9, the coupling parameter $c$ is gradually varied. It is seen that about the value of $c = 0.195$ the two multipliers cross the unit circle, implying a loss of stability.

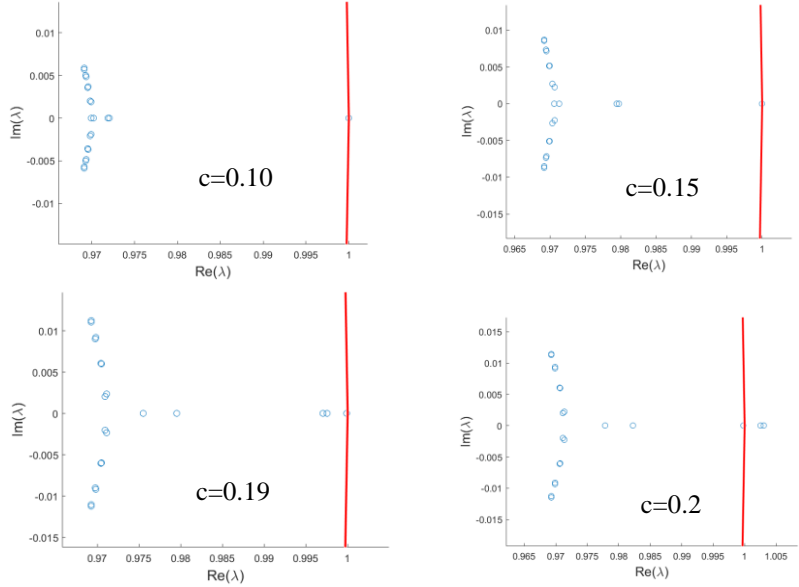

*Figure 9 – "Floquet multipliers" as a function of coupling $c$. Parameters: $a = 2.5$, $b = 0.5$, $\varepsilon = 0.01$*

### 2.2 Nucleation

A question arises what other possible attractors are stable within the system. Numerical simulations imply that the DB with a single particle within the LC, loses stability to DBs with a larger nucleus, say 3 or 5, prior to initiation of propagating front. The motivation is to conjecture that the propagating front becomes the lone stable solution, when all the possible DBs lose stability. It is possible, although somewhat more complicated, to study other breather configurations through the same procedure as outlined in 2.1-2.1.4. However, we present here a numerical study according to the following methodology. We instantiate DBs with different odd nucleus sizes (3, 5, 7, etc) in the "in-phase" mode of oscillation, by placing the assumed nucleus near the LC. Several examples are shown in

Figure 10. Then, we integrate the system numerically until the solution with the required nucleus size reaches steady state. Then, we gradually and slowly increase the bifurcation parameter (in this case $c$) to obtain the value of stability loss. The results for different nucleus size are shown in Figure 11. The graph implies that the value of $c_{cr}$ has a limiting value $\bar{c}_{cr} = 0.2011$ (obtained for nucleus of size 3). The value of $c_{cr}$ converges to a constant value from nucleus size of 5. These insights lead to a conclusion that the nucleation process occurs for a very narrow range of parameters. In this example it starts when the solution with nucleus size of 1 loses stability at $c = 0.1916$ and the process ends when the coupling reaches the value of $\bar{c}_{cr} = 0.2011$, where all possible nuclei lose stability.

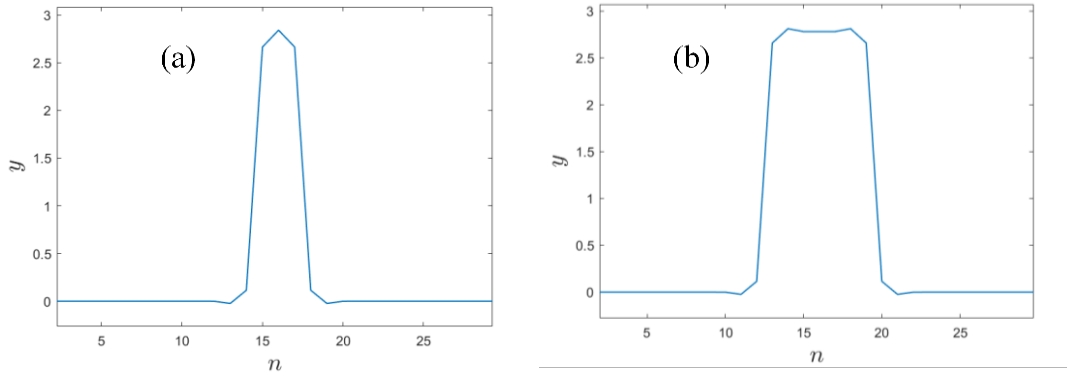

*Figure 10 – Discrete breathers with different nucleus size. a) Nucleus size: 3, b) Nucleus size: 7. Parameters: $c = 0.1$, $a = 2.5$, $b = 0.5$, $\varepsilon = 0.01$*

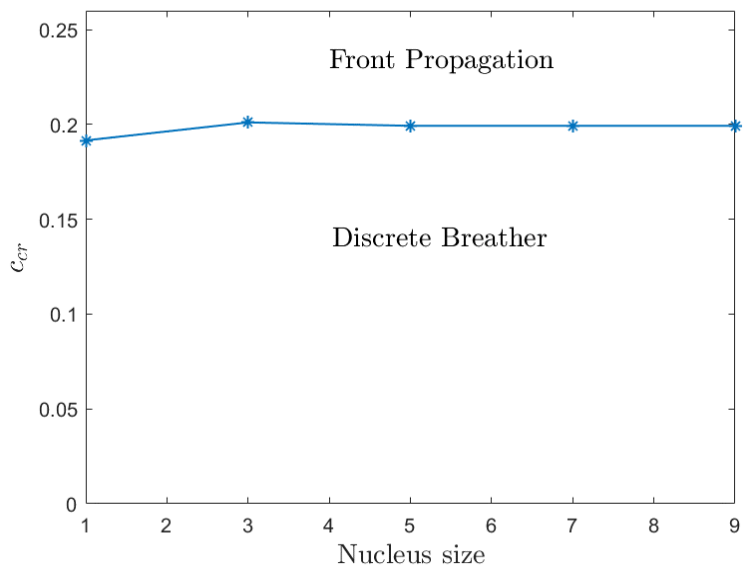

*Figure 11 – Critical coupling vs. nucleus size for "bulk" excitation. Parameters: $a = 2.5$, $b = 0.5$, $\varepsilon = 0.01$*

### 2.3 Travelling wave solutions

#### 2.3.1 Numerical solution of the VdP system (8)

The process of front propagation corresponds to successive transition of particles from the trivial equilibrium into the oscillation regime within the stable limit cycle. An example is shown in Figure 12.

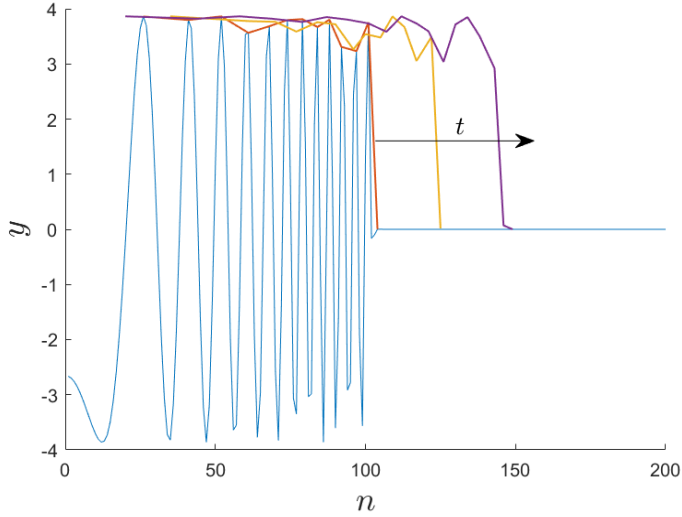

*Figure 12 – Front propagation in the couple VdP system. Parameters: $a = 4$, $b = 0.5, c = 0.6$, $\varepsilon = 0.01$, boundary conditions: open. The spatial form of the chain is plotted at three instances to demonstrate the steady propagation: $t = [14000, 17000, 20000]$.*

The next question to consider is how the front velocity is scaled with respect to the system parameters. We define a rescaled velocity $\hat{v} = \dfrac{v}{\varepsilon}$. First, we present the scaling of $\hat{v}$ with respect to coupling $c$ (Figure 13) which turns out to be asymptotically linear. This finding is similar to the friction-excited system. Second, we introduce parameter $\eta = \dfrac{b}{a^2}$, a dimensionless parameter, that represents the forcing in this system, and assume that the velocity can be scaled with respect to this single parameter. We show results obtained for different sets of $a, b$ in Figure 13, all collapsing on the same line, approving this presumed scaling.

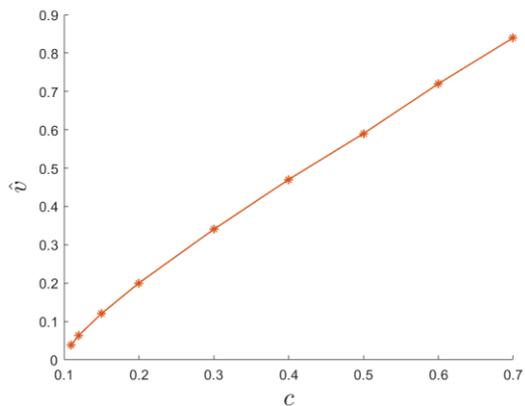

*Figure 13 – Scaling of $\hat{v}$ as a function of $c$ for the VdP model* (8)*; Parameters: $a = 4$, $b = 0.5$*

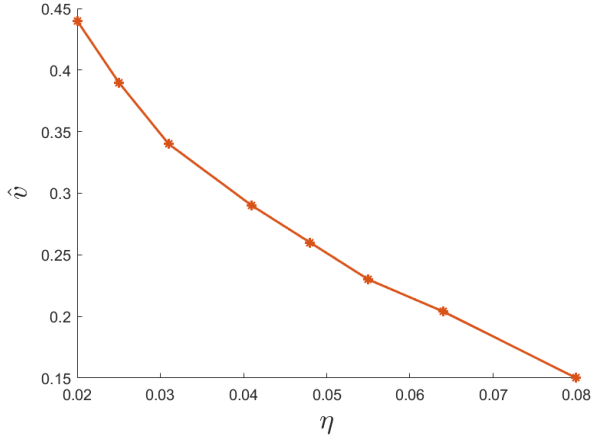

*Figure 14 – Scaling of $\hat{v}$ as a function of $\eta$ for the VdP model* (8)*; Parameters: $c = 0.3$*

In Figure 15 we show a map of the different dynamical states in the system, namely: front-propagation, discrete breathers and trivial fixed point. The data is shown in the space of $c - \eta$. Here $\eta$ is in a sense the equivalent forcing parameter parallel to $\tilde{v}_d$ in the friction-excited model. Quite remarkably, the map looks qualitatively similar to Figure 5 strengthening the assumed link between the two systems.

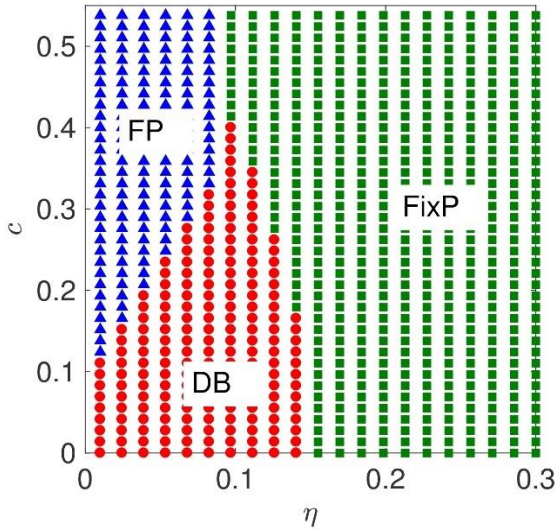

*Figure 15 – Map of the possible dynamical states of the VdP system* (8) *in the plane $(\eta, c)$. Front Propagation (FP), discrete breathers (DB) and fixed point (FixP) solutions correspond respectively to the blue triangles, red circles and green squares. Parameters used in simulations: 16 oscillators, $\varepsilon = 0.01$, initial conditions: displacement of the $16^{th}$ oscillator = 6, free-free boundary conditions.*

### 2.3.2 Analytical solution of a simplified model

In this subsection we suggest a rough model that, on the one hand, intends to keep the main qualitative features of the VdP model (8), and on the other hand, to allow an analytical solution for the front shape and velocity of propagation. We revisit the averaged equation of the model (34). Without affecting the generality, we set $\omega = 1$.

$$\dot{\varphi}_n - \frac{i\varepsilon c}{2}\left(2\varphi_n - \varphi_{n-1} - \varphi_{n+1}\right) + \varepsilon \frac{\varphi_n}{2}\left(1 - a\frac{|\varphi_n|^2}{4} + b\frac{|\varphi_n|^4}{8}\right) = 0 \qquad (43)$$

We simplify the equation by replacing the nonlinear term, with a piecewise linear approximation, thus separating the equation into a set of two linear equations (44). The straightforward advantage of this major simplification is that the obtained set is solvable analytically, while still preserving the bi-stable feature of the original model (8). The equilibrium values of Eq. (44) correspond to the steady state values at $\pm\infty$ of the original envelope of (8). Namely, the trivial equilibrium ($\varphi = 0$) and the stable limit cycle ($\varphi = \varphi_s$). The stitch between the equations, naturally, is set at the value of the unstable limit cycle ($\varphi = \varphi_{us}$). The equation somewhat resembles the Atkinson-Cabrera equations [26] of the coupled chain subjected to a piecewise parabolic on-site potential with equal curvatures.

$$\dot{\varphi}_n - \frac{i\varepsilon c}{2}\left(2\varphi_n - \varphi_{n-1} - \varphi_{n+1}\right) = -\frac{1}{2}\varepsilon \begin{cases} \varphi_n, & \varphi_n < \varphi_{us} \\ \varphi_n - \varphi_s, & \varphi_n > \varphi_{us} \end{cases} \qquad (44)$$

For convenience of further analysis, the force may be equivalently presented the following way:

$$F = -\frac{1}{2}\varepsilon\varphi_n + \frac{1}{2}\varepsilon\varphi_s \theta\left(\varphi_n - \varphi_{us}\right) \qquad (45)$$

Where $\theta\left(\varphi_n - \varphi_{us}\right)$ is a step-function, with discontinuity at $\varphi_n = \varphi_{us}$.

Let us assume that the system (44), (45) has a solution in the form of a travelling wave $\varphi_n = \varphi(n - vt)$ and introduce a new moving coordinate $\zeta = n - vt$:

$$\begin{aligned} \varphi_n &= \varphi(n - vt) = \varphi(\zeta) \\ \varphi_{n+1} &= \varphi(n - vt + 1) = \varphi(\zeta + 1) \\ \varphi_{n-1} &= \varphi(n - vt - 1) = \varphi(\zeta - 1) \end{aligned} \qquad (46)$$

We make another assumption, known in literature as an "admissibility condition" (47). Satisfaction of (47) means that the solution can be separated into two segments per range of $\zeta$. For, $\zeta < 0$ the entire solution satisfies $\varphi > \varphi_{us}$, whereas for $\zeta > 0$ it satisfies $\varphi < \varphi_{us}$. This condition doesn't restrict the scope of the analysis as far as the self-consistency is satisfied. The obtained solutions should be checked for obeying to this condition, otherwise they should be removed due to inconsistency.

$$\varphi_n > \varphi_{us} \Leftrightarrow \zeta < 0 \qquad (47)$$

Let us define $v \equiv \varepsilon\hat{v}$. Substitution of (46), (45), (47) into (44) yields the following complex advance delay equation:

$$-\hat{v}\dot{\varphi}(\zeta) - \frac{ic}{2}\left[2\varphi(\zeta) - \varphi(\zeta + 1) - \varphi(\zeta - 1)\right] + \frac{1}{2}\varphi = \frac{1}{2}\varphi_s \theta(-\zeta) \qquad (48)$$

Equation (48) is solved analytically by applying the Fourier transform, similarly to previous works [26, 27, 28, 29, 30]. In particular, in [31] a system with environmental viscosity was studied and this led to model that is similar to the current one in a sense that the dispersion relation was complex. We apply the same technique here and the solution is written in the following form:

$$\varphi(\zeta) = \begin{cases} \varphi_s + \dfrac{1}{2}\varphi_s \sum_{k \in M^-} \dfrac{e^{ik\zeta}}{kL_k} & \zeta < 0 \\ -\dfrac{1}{2}\varphi_s \sum_{k \in M^+} \dfrac{e^{ik\zeta}}{kL_k} & \zeta > 0 \end{cases} \tag{49}$$

Where the dispersion relation of (48) and its derivative with respect to $k$ are given by:

$$L(k) = -i\hat{v}k - 2ic\sin^2\dfrac{k}{2} + \dfrac{1}{2} \tag{50}$$

$$L_k(k) = -\dfrac{1}{2k} + \dfrac{2ic}{k}\sin^2\dfrac{k}{2} - ic\sin k \tag{51}$$

The roots of the dispersion relation (50) are denoted as:

$$\begin{aligned} M^+ &= \{k : L(k) = 0,\ \mathrm{Im}\,k > 0\} \\ M^- &= \{k : L(k) = 0,\ \mathrm{Im}\,k < 0\} \end{aligned} \tag{52}$$

In order to obtain an expression for velocity as a function of other parameters (kinetic relation), we use the boundary condition $\varphi(0) = \varphi_{us}$, and obtain:

$$\sum_{k \in M^+} \dfrac{1}{kL_k} = -\dfrac{2\varphi_{us}}{\varphi_s} \tag{53}$$

By taking an absolute value on (53) and substituting the known values for $|\varphi_{us}|, |\varphi_s|$, one obtains:

$$\left|\sum_{k \in M^+} \dfrac{1}{kL_k}\right| = \dfrac{2|\varphi_{us}|}{|\varphi_s|} = 2\sqrt{\dfrac{1-\sqrt{1-8\eta}}{1+\sqrt{1-8\eta}}},\quad \eta \equiv \dfrac{b}{a^2} \tag{54}$$

Eq. (54) is an implicit expression for the velocity of propagation $\hat{v}$ as a function of $\eta$, $c$. $\hat{v}$ can't be explicitly extracted due the infinite number of complex roots $M^+$. It can be solved numerically to find the value of $\hat{v}$ for any set of parameters $\eta$, $c$. An example of the solution for $\varphi$ is presented in Figure 16. The solution clearly satisfies the admissibility condition and converges to constant values at $\pm\infty$ of $\varphi(\zeta \to \infty) = 0,\ \varphi(\zeta \to -\infty) = \varphi_s$.

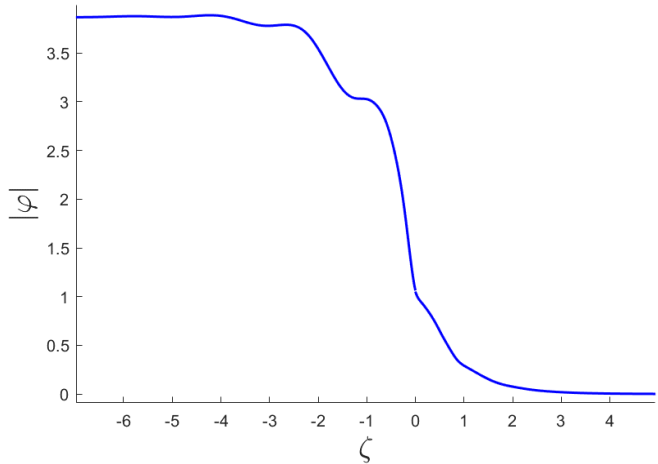

*Figure 16 – Analytical solution for $|\varphi|$ as a function of $\zeta$ for parameters: $a = 4, b = 0.5, c = 0.5$.*

Due to the two major simplifications of averaging the original equations and substituting the nonlinear term in (43) with a piecewise linear approximation, we don't expect a quantitative match of solution (49), (54) and original model (8). Instead, we focus on two qualitative behaviors. First, we are interested in the scaling of $\hat{v} = \hat{v}(c)$ shown in Figure 17. The scaling is asymptotically linear, thus recovering one of the major features of model (8). The second scaling we show is of $\hat{v}$ with respect to parameter $\eta$ that represents the forcing in the system (Figure 18). The qualitative behavior is similar to the one observed for the VdP system. These two similarities imply the equivalence of the two models in terms of dynamical behavior.

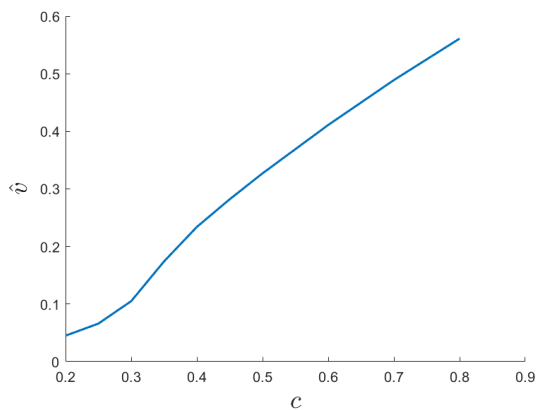

*Figure 17 – Scaling of $\hat{v}$ as a function of $c$ for the simplified model (43); Parameters: $a = 4, b = 0.5$.*

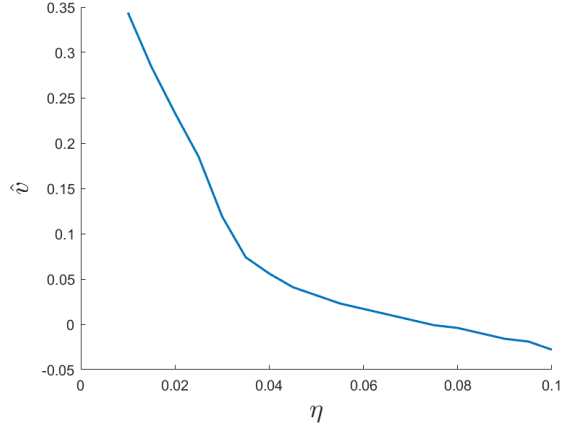

*Figure 18 – Scaling of $\hat{v}$ as a function of $\eta$ for the simplified model* (43)*; Parameters: $c = 0.3$*

3. Conclusions

Frictional interfaces are well-know to exhibit a rich variety of dynamical phenomena due to friction. In this paper we have tried to shed some light on the transition between spatially *localized* and *propagating* stick-slip motions. To this end we have studied the dynamical behavior of two dynamical models: (model I) a friction-excited chain of weakly coupled oscillators excited by a frictional moving belt and (model II) a chain of Van der Pol oscillators. We have shown that, even if the two models are different, their dynamical behavior share very similar features. In both models, discrete breathers (i.e. spatially localized periodic solutions) are observed for low coupling. When the elastic coupling is increased, the discrete breather solution loses stability and high amplitude limit cycles start to propagate and spread within the chain. For both models the transition from discrete breathers to propagating limit cycles happens through a nucleation process. When a critical coupling stiffness is reached the limit cycles start to propagate in the chain with a velocity that scales linearly with the elastic coupling coefficient. We suggest that the similarities between the two models originates from the similar dynamical behavior of the unit cell: the "isolated oscillator", in both models, has a fixed point surrounded by two limit cycles, the inner one being unstable and the outer one being stable. These results imply on a possible robustness of models with similar topology as were addressed in this work, Hence, dynamical features similar to what was demonstrated in this work are expected for models that are different quantitively but preserve the same single-unit topology.

**Author Contribution Statement**: All authors conceived the work. IBS and OVG developed the model for the VdP chain. AP and NH developed the model for the friction-excited chain. IBS and AP ran the simulations, wrote the manuscript and created the figures. All authors revised the manuscript up to its final version.

**Acknowledgements**: AP is thankful to the DFG (German Research Foundation) for funding the project PA 3303/1-1. IBS and OVG are grateful to Israel Science Foundation (grant 1696/17) for financial support.